\documentclass{kluwer}
\usepackage[dvips]{graphicx}

\newcommand\bgreek[1]{ \mathchoice
    {\hbox{\boldmath$\displaystyle{#1}$\unboldmath}}%
    {\hbox{\boldmath$\textstyle{#1}$\unboldmath}}%
    {\hbox{\boldmath$\scriptstyle{#1}$\unboldmath}}%
    {\hbox{\boldmath$\scriptscriptstyle{#1}$\unboldmath}}}

\newcommand\la{\;
  \raise0.3ex\hbox{$<$\kern-0.75em\raise-1.1ex\hbox{$\sim$
  }}\;\hskip-2pt }
\newcommand\ga{\;
  \raise0.3ex\hbox{$>$\kern-0.75em\raise-1.1ex\hbox{$\sim$
  }}\;\hskip-2pt }

\newcommand{\cmcube}{\,{\rm cm^{-3}}}

\newcommand{\g}{\,{\rm g}}

\newcommand{\kms}{\,{\rm km\,s^{-1}}}

\newcommand{\K}{\,{\rm K}}

\newcommand{\AU}{\,{\rm AU}}

\newcommand{\Fig}[1]{Fig.~\ref{#1}}
\newcommand{\EQ}{\begin{equation}}
\newcommand{\EN}{\end{equation}}

\newcommand{\nab}{\bgreek{\nabla}}

\newcommand{\uuu}{\mbox{\boldmath $u$} {}}

\newcommand{\ymn}[3]{ #1, {MNRAS,} {#2}, #3}
\newcommand{\yana}[3]{ #1, {A\&A,} {#2}, #3}
\newcommand{\yapj}[3]{ #1, {ApJ,} {#2}, #3}
\newcommand{\ynat}[3]{ #1, {Nat,} {#2}, #3}
\newcommand{\ysph}[3]{ #1, {Sol. Phys.,} {#2}, #3}
\newcommand{\ypasj}[3]{ #1, {Publ. Astron. Soc. Japan,} {#2}, #3}
\newcommand{\yproc}[5]{ #1, in {#3}, ed. #4 (#5), #2}

\begin{document}
\begin{article}
\begin{opening}

\title{Outflows from dynamo-active protostellar accretion discs}
\author{Brigitta \surname{von Rekowski}}
\institute{Department of Astronomy \& Space Physics,
  Uppsala University, Box 515, 751 20 Uppsala, Sweden}
\author{Axel \surname{Brandenburg}}
\institute{NORDITA, Blegdamsvej 17, DK-2100 Copenhagen \O, Denmark}
\author{Wolfgang \surname{Dobler}}
\institute{Kiepenheuer-Institut f\"ur Sonnenphysik,
  Sch\"oneckstr. 6, 79104 Freiburg, Germany}
\author{Anvar \surname{Shukurov}}
\institute{School of Mathematics \& Statistics, University of
Newcastle, Newcastle NE1 7RU, U.K.}

\runningauthor{B.\ von Rekowski et al.}
\runningtitle{Outflows from dynamo-active protostellar accretion discs}
\date{\today}

\begin{abstract}
An axisymmetric model of a cool, dynamo-active accretion disc is applied to
protostellar discs.
Thermally and magnetically driven outflows develop
that are not collimated within $0.1 \AU$.
In the presence of a central magnetic field from the protostar,
accretion onto the protostar is highly episodic, which is in
agreement with earlier work.
\end{abstract}

\maketitle

\section{Introduction}

Outflows from protostellar discs have been observed in
many star-forming regions.
These outflows are generally believed to be driven by magneto-centrifugal
acceleration from the accretion disc around the protostar.
Yet, most theoretical approaches to understanding stellar
outflows have ignored or at least greatly oversimplified
both accretion disc physics and magnetic field evolution.
Instead, most effort so far has focused around the physics
of magnetised outflows and the collimation process
assuming a fixed magnetic field.
One of the perhaps most convincing simulations of the collimation of
outflows have come from Ouyed et al.\ (1997) and Ouyed \& Pudritz (1997)
who considered a disc corona with an imposed field aligned with
the rotation axis of the
disc and modelled the disc physics as a boundary condition.

In recent years several groups have begun to approach the problem
from different directions.
One approach is to start from accretion disc physics and to
extend the simulation domain in the vertical direction, so as
to capture parts of the outflow acceleration region.
The global simulations of Hawley (2000)
are among the most impressive
attempts to treat the problem as a whole.
Here the magnetic field is generated by the disc dynamo driven by the magneto-rotational
instability and some form of
outflow is indeed seen in the vicinity of the rotation axis.
However, the dynamo still seems to be unable to produce a large
scale poloidal field (which seems to be required for driving
systematic outflows), and
thermodynamics is not included
to allow the disc to cool sufficiently to become
geometrically thin.
Other groups have focused on the dynamo aspect and have used
thin disc theory as input to their models
(Reyes-Ruiz \& Stepinski 1995)
or have studied the effect of vertical outflows on the
disc dynamo (Bardou et al.\ 2001).

In the present paper we review the approach of von Rekowski et al.\ (2003)
which is conceptually closest to the dynamo approach mentioned above.
The effect
of the Lorentz force on the gas motions is taken into account.
For a thin accretion disc, it is required that
near-Keplerian shear velocity be the result
of a balance mainly between centrifugal force and gravity.
This is only possible if the disc is cool, and remains cool notwithstanding
the adiabatic heating associated with the concentration
of gas near the midplane by the vertical component of gravity
during the disc formation.

In order to stay conceptually as close as possible to the polytropic
setup of Ouyed \& Pudritz (1997) we adopt a bi-polytropic model
with low entropy in the disc and high entropy in the exterior
which we refer to as halo or corona (see the next section).

Since we restrict ourselves to two-dimensional geometry, we
model the disc dynamo using mean-field ($\alpha\Omega$ dynamo) theory.
The applicability of this theory to protostellar accretion discs, where the
turbulence (and the dynamo) is driven by the magneto-rotational (or Balbus-Hawley)
instability, will also be discussed in the next section.

\section{Modelling a cool dynamo-active accretion disc}

Like Ouyed \& Pudritz (1997) we start with an equilibrium
corona, where we assume constant entropy and hydrostatic equilibrium
according to $c_p T(r)=GM_*/r$.
(Here, $T$ is temperature, $G$ is the gravitational constant,
$M_*$ is the mass of the protostar, $r$ is the spherical radius,
and $c_p$ is the specific heat at constant pressure.)
In order to make the disc cooler, we prescribe a geometrical
region for the disc (see \Fig{Fig-structure}) and choose an entropy
contrast between disc and corona such that the initial disc temperature
is about $3\times10^3 \K$ in the bulk of the disc.
Given the degree of simplification of our model, we have refrained
from modelling the
flaring of the disc.
The low disc temperature corresponds to a high disc density of about
$10^{-10} \dots 10^{-9} \g \cmcube$.

\begin{figure}
  \centering
  \includegraphics[height=4cm]{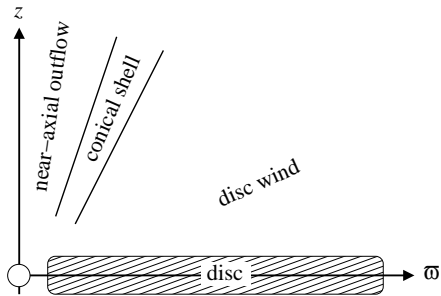}
  \caption{General structure of the outflows typically obtained in our
    model.
    The cool, dense disc emits
    (a) a thermally driven wind
    (slow, hot, dense, magnetised, rotating),
    (b) a magneto-centrifugally driven outflow in a conical shell
    (faster, cooler, less dense, magnetised, quickly rotating),
    and
    (c) a thermally driven outflow near the axis
    (slow, hot, dense, weakly magnetised and weakly rotating).
  }
  \label{Fig-structure}
\end{figure}

Hydrostatic equilibrium requires $(\uuu\cdot\nab)\uuu
=-(1/\rho)\nab p-\nab\Phi$, where $\uuu=(0,u_\varphi,0)$ is the velocity field,
$p$ is gas pressure, $\rho$ is gas density, and $\Phi$ is the gravitational
potential.
Since specific entropy $s$ is given, it is convenient to express
$(1/\rho)\nab p$ in terms of $s$ and specific enthalpy $h$, using the
relation $\nab h=(1/\rho)\nab p+T\nab s$ (first law of thermodynamics)
and $h=c_p T$.
In this way we obtain $h$ by integrating the vertical balance equation,
\EQ
-\partial(h+\Phi)/\partial z + h\,\partial(s/c_p)/\partial z = 0,
\EN
from large $z$ (where $h+\Phi=0$) down to $z=0$.
The initial rotation velocity, $u_{\varphi0}$, follows from the radial
balance equation,
\EQ
-u_{\varphi0}^2/\varpi = -\partial(h+\Phi)/\partial\varpi
+ h\,\partial(s/c_p)/\partial\varpi.
\EN
In the disc, $h=c_p T$ is small, so $u_{\varphi0}$ is close to the
Keplerian velocity, while the corona does not rotate initially,
and is supported by the pressure gradient.
Within the disc, we adjust density with a forcing term, so that a similar
disc structure is maintained on reasonable time scales.

For the disc dynamo, the most important parameter is the
dynamo coefficient $\alpha$ in the mean-field induction equation.
The $\alpha$ effect is antisymmetric about the midplane, restricted to the disc,
and roughly
a negative multiple of the kinetic helicity, so $\alpha$ is normally
positive in the upper disc half.
However, there is evidence that in accretion discs, where
turbulence is driven by the magneto-rotational instability, $\alpha$
is negative in the upper disc half (Brandenburg
et al.\ 1995, Ziegler \& R\"udiger 2000).
A simple theoretical explanation is given in Brandenburg (1998), and
a similar result has been obtained by R\"udiger \& Pipin (2000) who
considered magnetically driven turbulence; see also
R\"udiger et al.\ (2001).

Is there any evidence that the $\alpha\Omega$ dynamo picture
is correct and applicable to accretion discs?
One piece of evidence
is that the specific dependence of the large scale field geometry on
the boundary conditions is the same both in simulations of shearing sheet
turbulence driven by the magneto-rotational instability and in
$\alpha\Omega$ dynamo models in the same geometry.
Since the dynamo-generated magnetic field is of large scale and since it
extends to the boundaries, one must expect a detailed dependence of its
geometry on the boundary conditions.
In discs with negative $\alpha$ in the upper half,
with (pseudo-)vacuum boundary conditions one gets an oscillatory large scale field
whose horizontal components are symmetric about the midplane.
With perfect-conductor or finite-conductivity boundary conditions,
one gets a non-oscillatory field whose horizontal components are
antisymmetric about the midplane.
The {\it same} is seen both in three-dimensional simulations and in
solutions of the $\alpha\Omega$ dynamo problem; see Brandenburg (1998).

Bardou et al.\ (2001) found that vertical outflows do not change
the symmetry or temporal behaviour of the dynamo-generated magnetic field.
However, especially in the linear case for negative $\alpha$ in the upper half and
finite-conductivity boundary conditions,
a strong outflow tends to switch the dynamo off unless the dynamo
number is large enough.

\begin{figure}
 \centering
 \includegraphics[height=6cm]{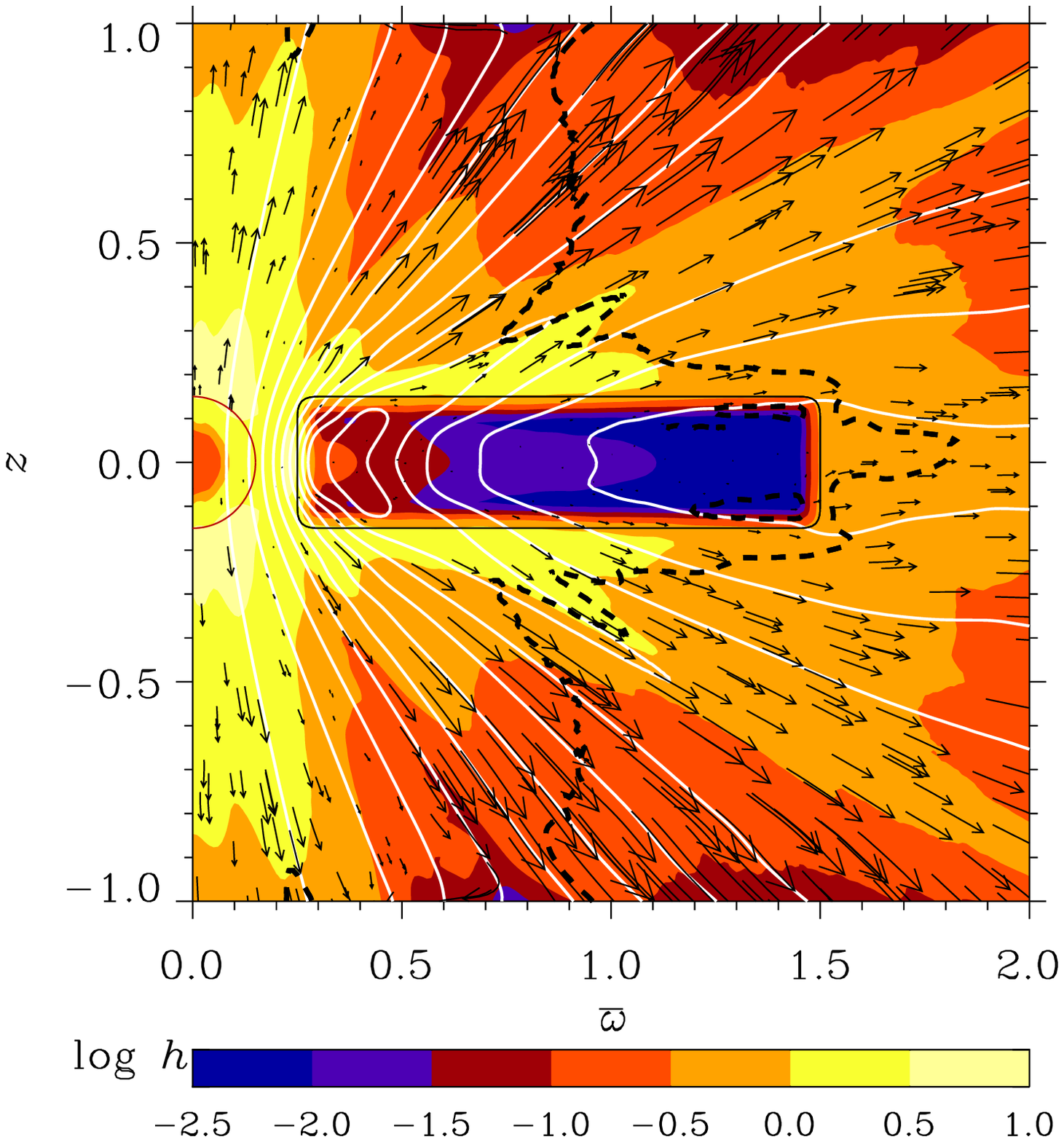}
 \includegraphics[height=6cm]{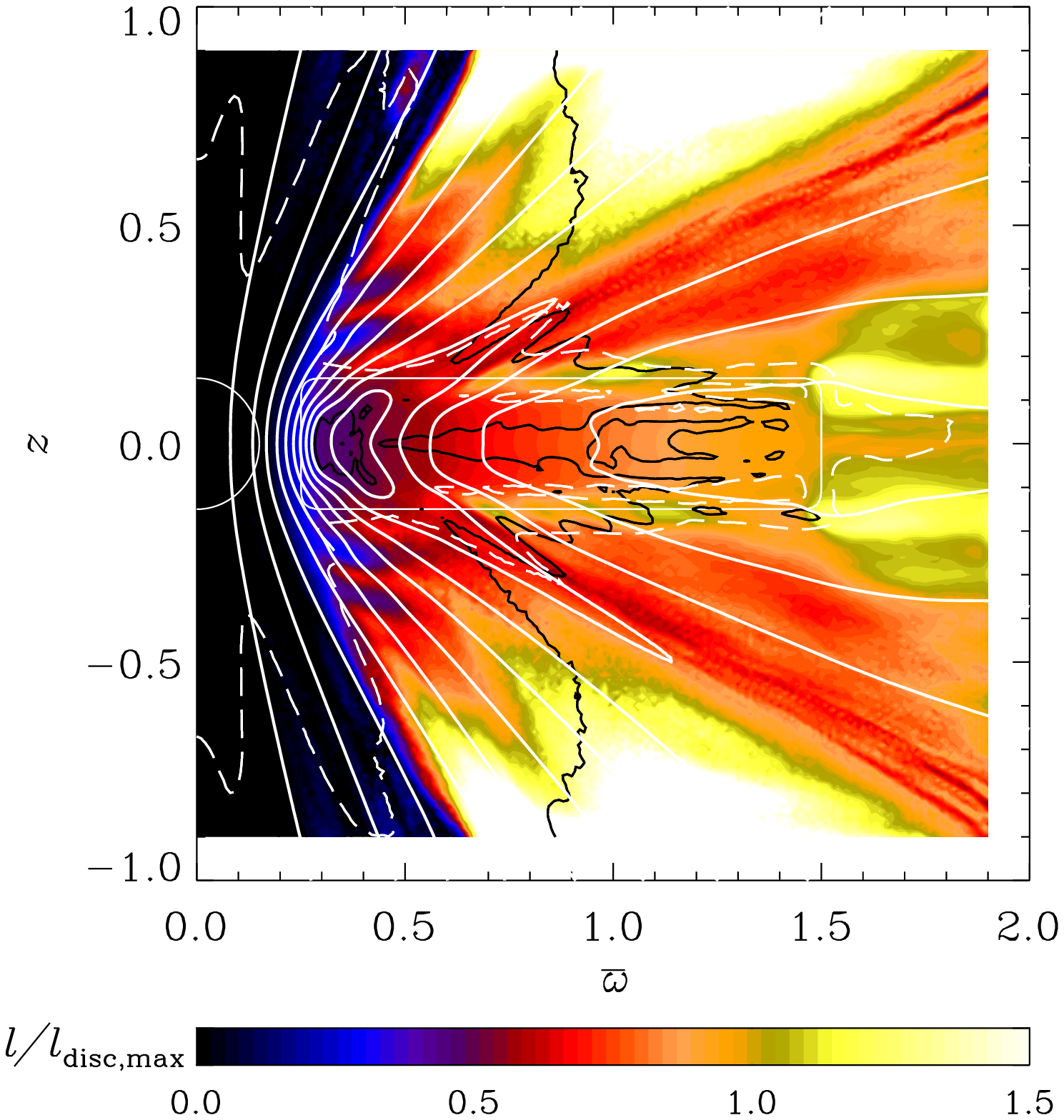}
 \caption{
  Left:
  poloidal velocity vectors and poloidal magnetic field lines (white)
  superimposed on a colour scale representation of $\log_{10}h$. Specific
  enthalpy $h$ is directly proportional to temperature $T$, and
  $\log_{10}h=(-2,-1,0,1)$ corresponds to $T\approx
  (3{\times}10^{3},3{\times}10^{4},3{\times}10^{5},3{\times}10^{6})\,\mbox{K}$.
  The black dashed line shows the fast magnetosonic surface.
  The disc boundary is shown
  with a thin black line, the stellar surface is marked in red.
  The dynamo $\alpha$ coefficient is negative in the upper disc half,
  resulting in roughly dipolar magnetic
  symmetry. Averaged over times $t\approx 897 \dots 906\,{\rm days}$.
  Right:
  colour scale representation of the specific
  angular momentum, normalised by the maximum angular momentum in the disc,
  with poloidal magnetic field lines superimposed (white).
  The black solid line shows the Alfv\'en surface, the white dashed
  line the sonic surface. Same model as left and averaged over same times.
 }
 \label{Fig1}
\end{figure}

\section{Structured outflow and acceleration mechanism}

Figure~\ref{Fig1}, left panel, illustrates that an outflow develops
that has a well-pronounced structure. Within a conical shell originating
from the inner edge of the disc, the terminal outflow speed exceeds
$500\kms$, and temperature and density are lower than elsewhere.
The inner cone around the axis is the hottest and densest region
where the stellar wind speed reaches about $150\kms$. The wind that develops
from the outer parts of the disc has intermediate values of the speed
(see the sketch in \Fig{Fig-structure}).

The structured outflow is driven by a combination of different processes.
A significant amount of angular momentum is transported outwards
from the disc into the wind along magnetic field lines, especially along the
strong lines within the conical shell (see \Fig{Fig1}, right panel).
The magnetic field geometry is such that the angle between the rotation
axis and the field lines threading the disc exceeds $30^\circ$ at the disc surface,
which is favourable for magneto-centrifugal acceleration (Blandford \& Payne
1982). However, the Alfv\'en surface is so close to the disc surface
at the outer parts of the disc that acceleration there is mainly due to
the pressure force. In the conical shell, however, the outflow is highly
supersonic but sub-Alfv\'enic, with the Alfv\'en radius a few times larger
than the radius at the footpoint of the field lines at the disc surface.
The lever arm of about 3 is
sufficient for magneto-centrifugal acceleration to dominate (cf.\ Krasnopolsky
et al.\ 1999). The inner cone has very low angular momentum and the magnetic lines are
inclined to the axis by less than $30^\circ$ so that the stellar wind can only be
pressure driven. This picture is confirmed by looking at the ratio of
magneto-centrifugal to pressure forces, which is much larger than unity
in the conical shell. This is also the region where magnetic pressure
produced by the toroidal field exceeds gas pressure,
indicating that the outflow in the conical shell is confined by the toroidal field.

\section{Magnetic coupling between star and disc}

\begin{figure}
 \centering
 \includegraphics[height=5.5cm]{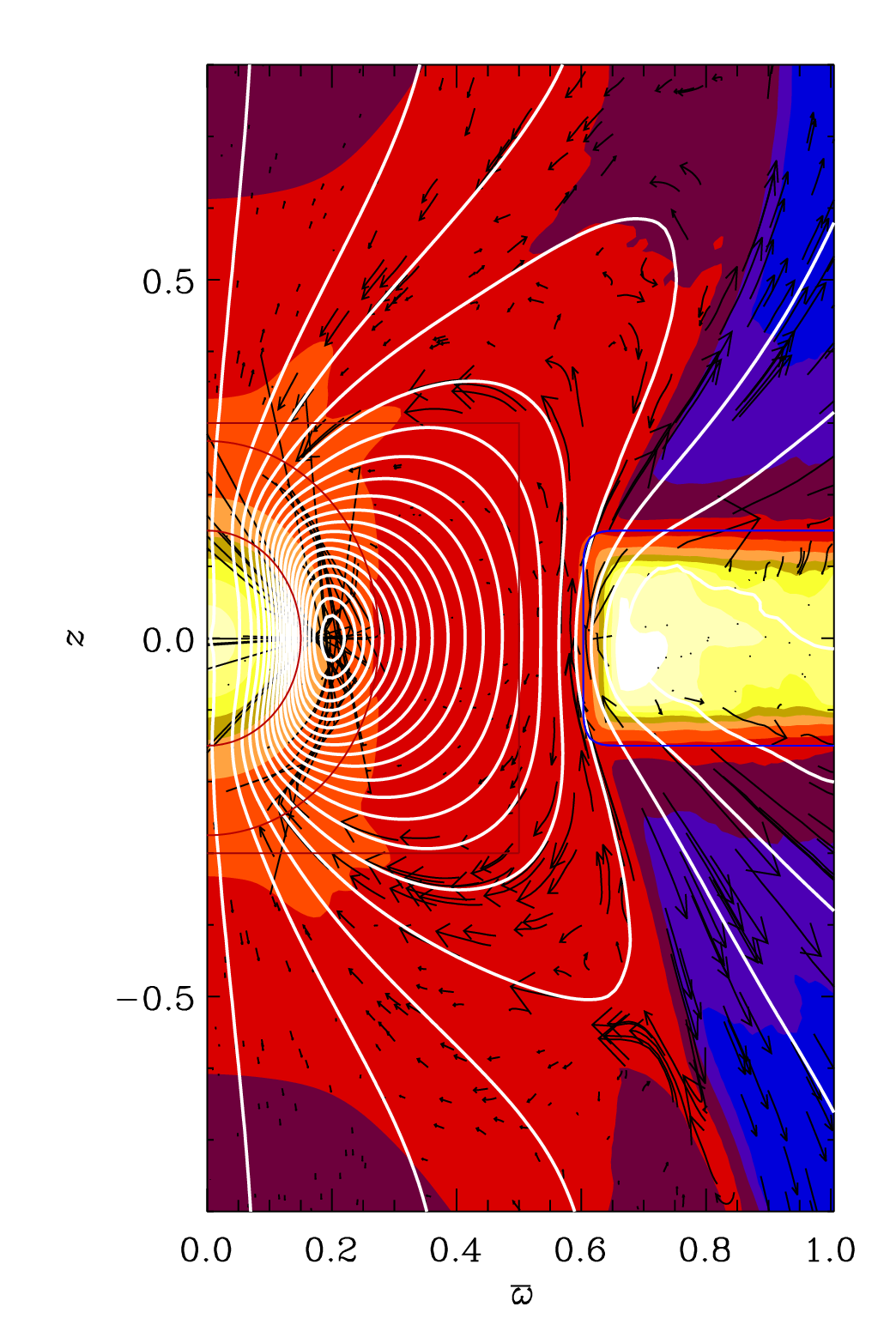}
 \includegraphics[height=5.5cm]{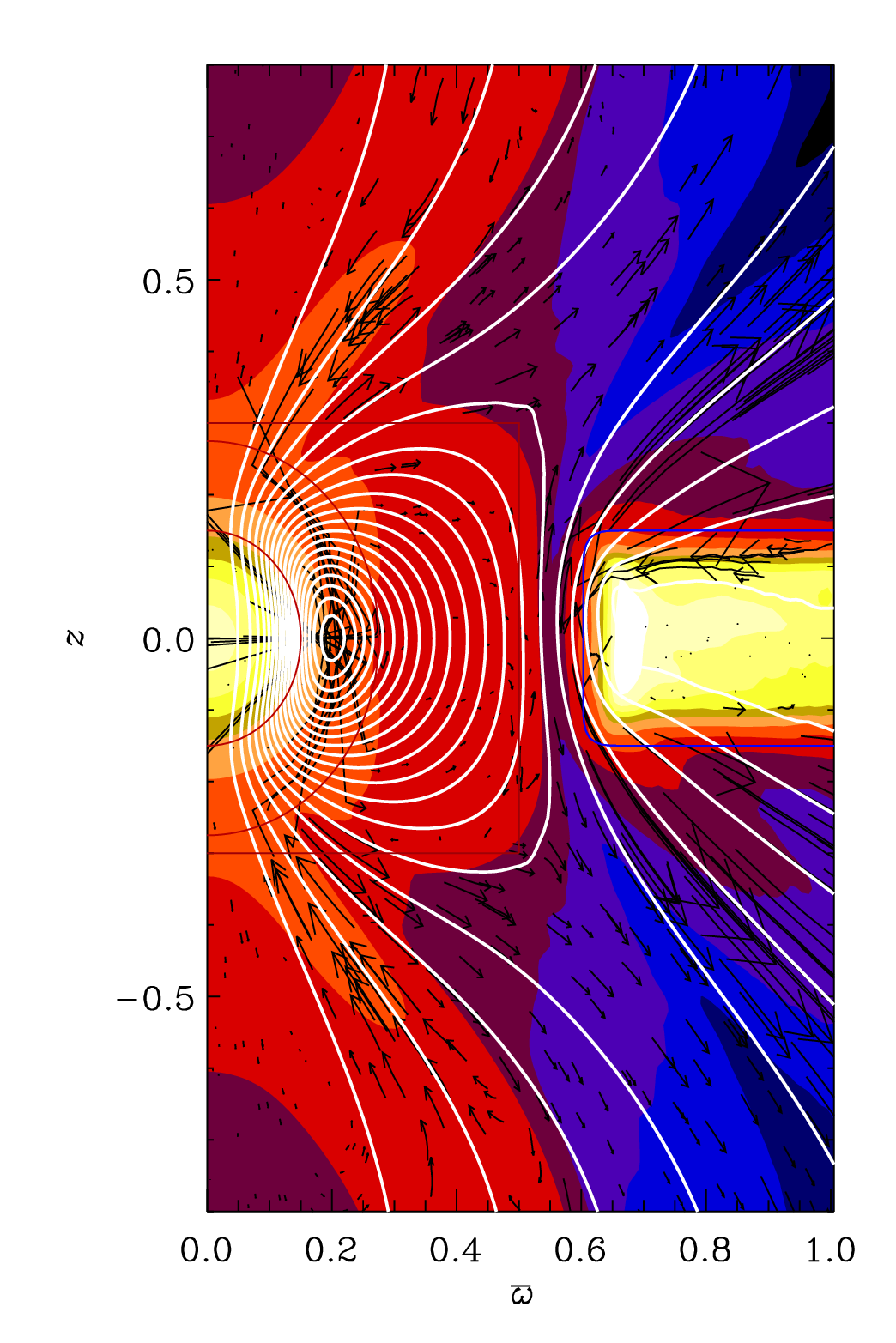}
 \caption{
  Colour scale representation of the density (bright colours indicate high values;
  dark colours, low values) with poloidal magnetic field lines superimposed (white)
  and the azimuthally integrated mass flux density shown with arrows (not shown in the disc).
  Left: at a time $t\approx 150\, {\rm days}$, when the star is magnetically connected to the disc.
  Right: at a time $t\approx 156\, {\rm days}$, when the star is disconnected from
  the disc.
 }
 \label{Fig2}
\end{figure}

Observations of T~Tauri stars suggest that the stellar spin-down can be connected
with a circumstellar accretion disc. Magnetic fields are believed to play an
important r\^ole in the interaction between the star and its surrounding disc,
resulting in particular in a spin-down of the star due to magnetic braking.
Indeed, magnetic fields as strong as $10^3$~Gauss have been
detected on T~Tauri stars, and there is evidence for hot and cool spots on the
stellar surface.
Although the structure of the stellar magnetosphere is yet unknown, analytical
models have been developed by Hartmann et al.\ (1994) and Shu et al.\ (1994)
that assume that a stellar dipole threads a surrounding disc. In these
static magnetospheric accretion models, disc matter is channelled along
magnetospheric field lines to form a funnel-shaped polar accretion flow.

More realistic time dependent numerical simulations by Hirose et al.\ (1997)
and Goodson et al.\ (1997) show episodic reconnection behaviour between a stellar
dipolar magnetosphere and an imposed disc field, resulting in episodic mass
transfer from the disc to the star. These authors find inner stellar jets and outer
disc winds that are both driven by magneto-centrifugal processes.

Figure~\ref{Fig2} shows our model with disc dynamo, where we add a stellar dipole
that is anchored within the star, but can evolve with time outside the star
(von Rekowski \& Brandenburg 2003).
For the strong surface stellar field of about $6\,{\rm kG}$ considered here,
the time dependent accretion is correlated with the magnetic star--disc coupling.
When the star is connected to the disc, accretion is along field lines (see
\Fig{Fig2}, left panel). When the star is disconnected from the disc, an enhanced
wind carries away excess angular momentum (see \Fig{Fig2}, right panel).
The sum of magnetic and accretion torques in the outward direction is positive
at radii between star and disc, which means a stellar spin-down.

\begin{figure}
 \begin{center}
 \includegraphics[height=4.125cm]{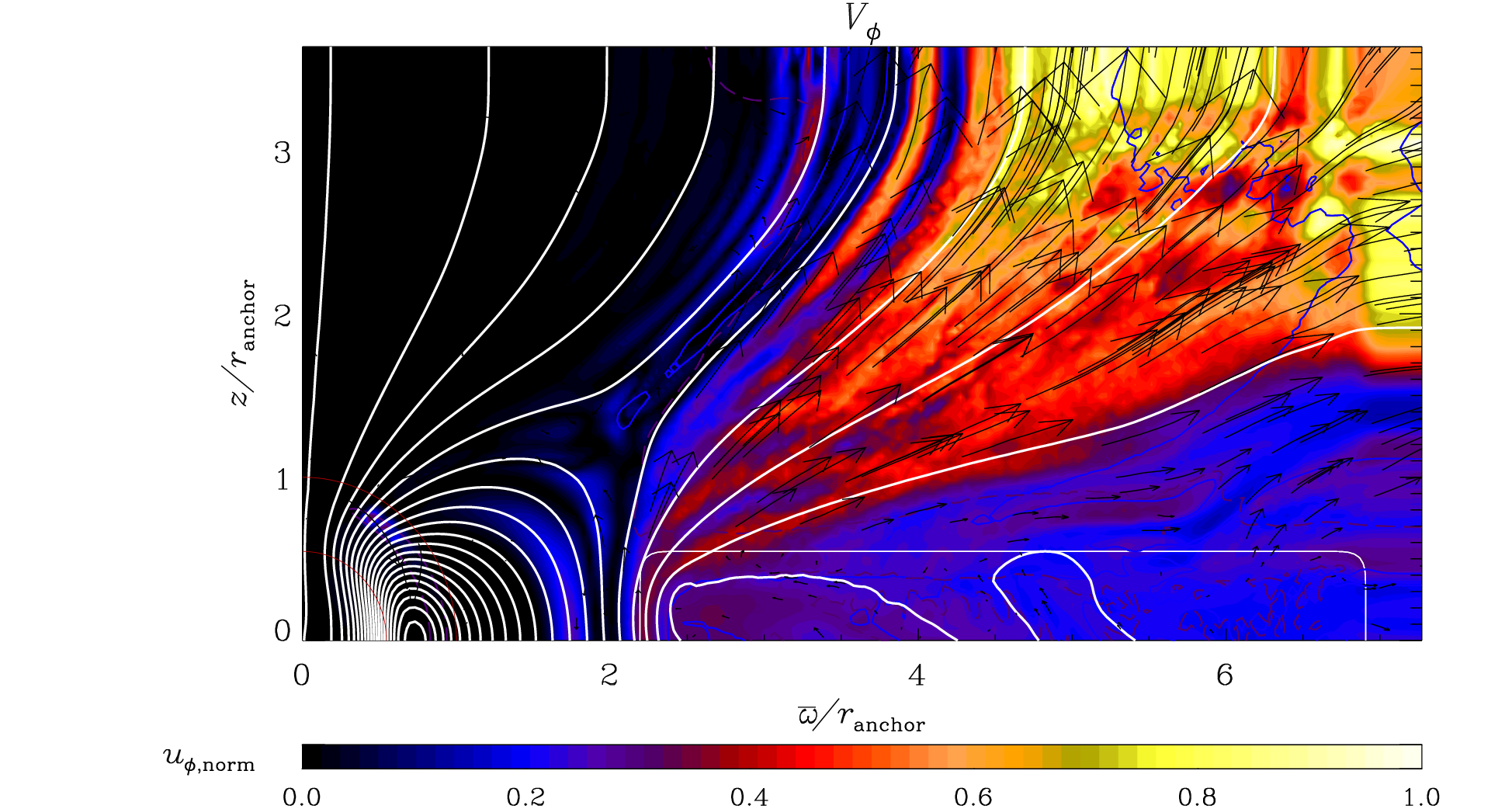}
 \includegraphics[height=4.25cm]{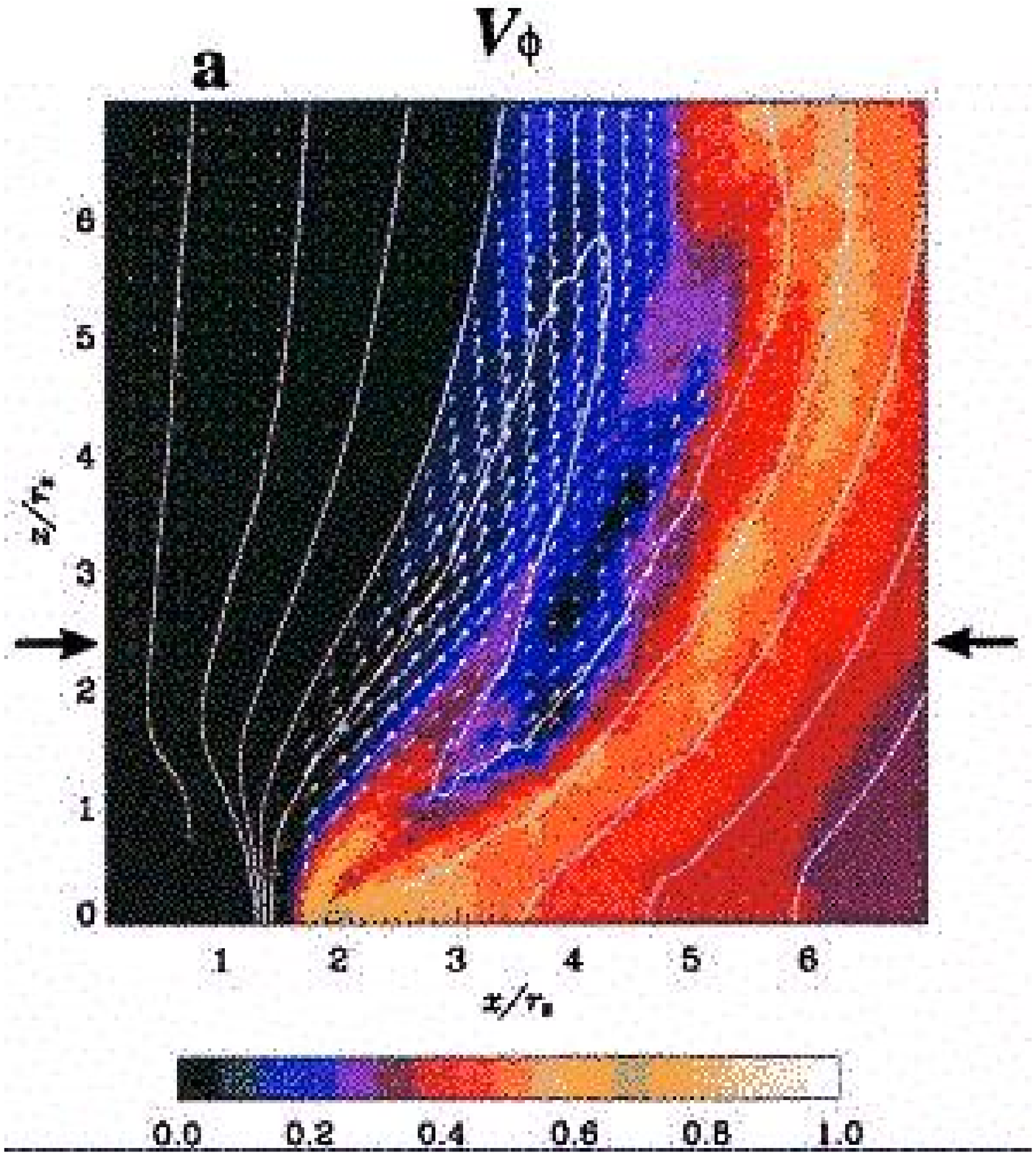}
 \end{center}
 \begin{center}
 \includegraphics[height=4.125cm]{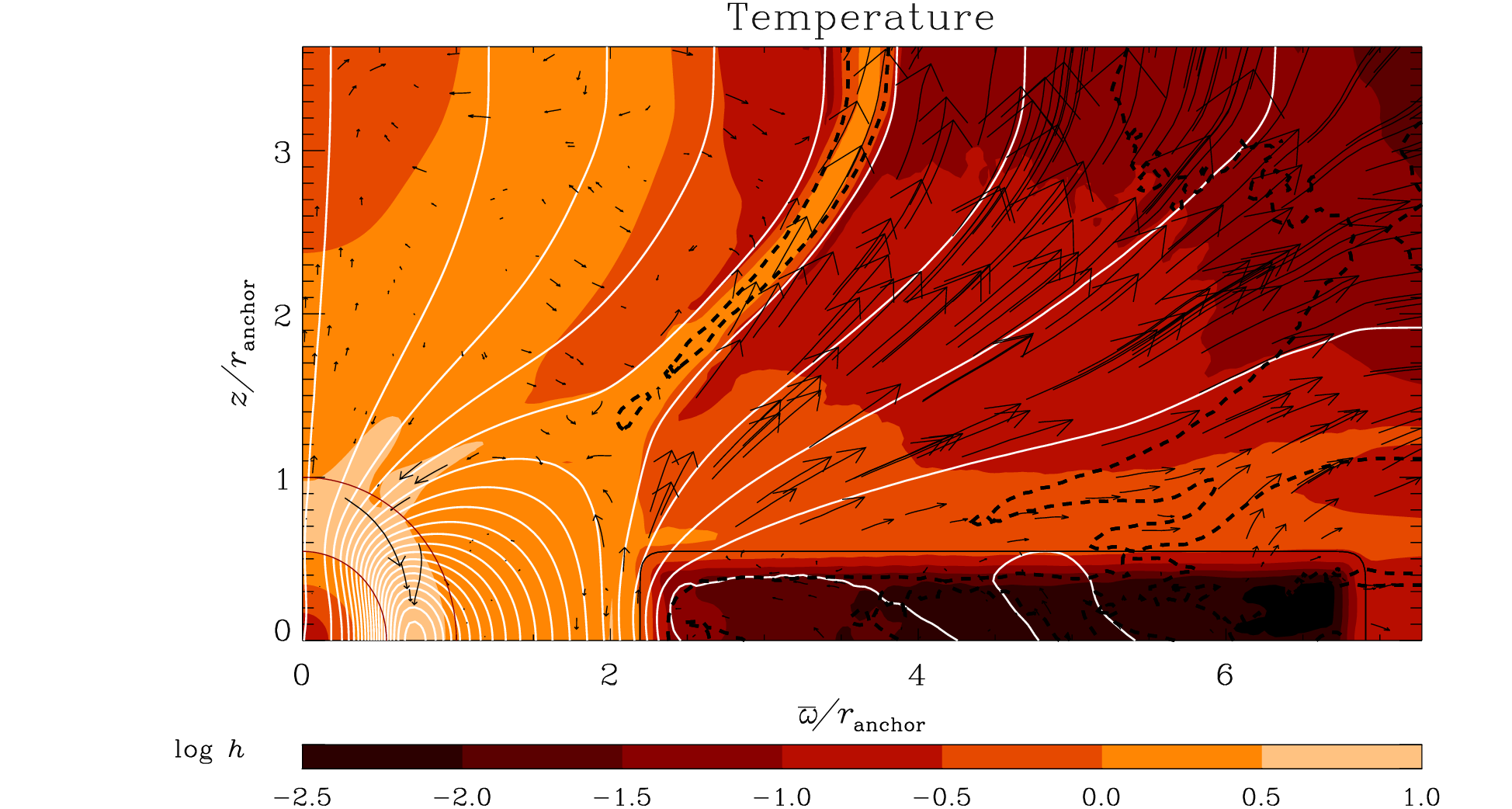}
 \includegraphics[height=4.25cm]{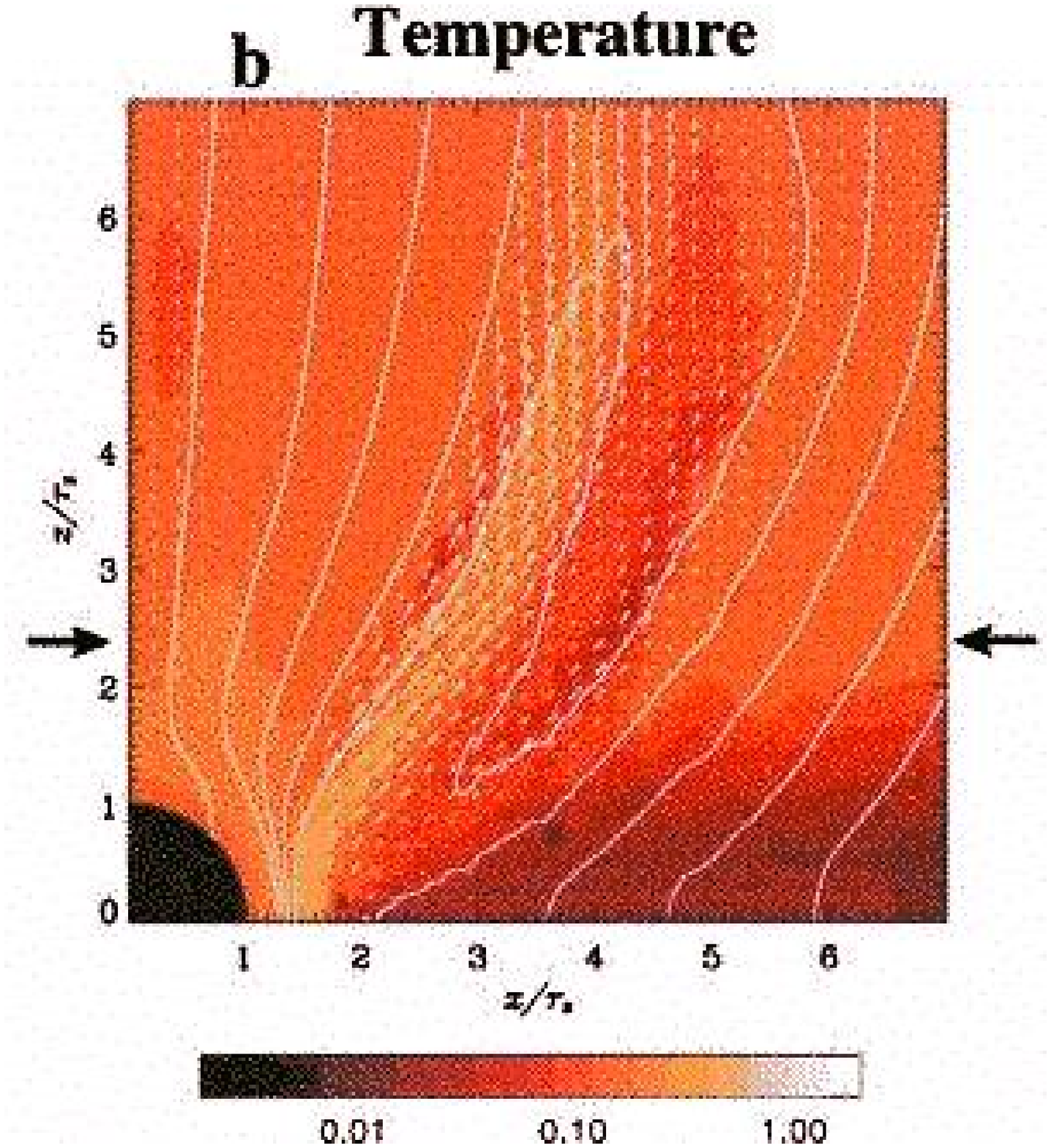}
 \end{center}
 \caption{
  Colour scale representation of the normalised azimuthal velocity (top) and
  temperature (bottom) with poloidal magnetic field lines (white)
  and poloidal velocity vectors superimposed.
  Left: simulations of outflows from protostellar accretion discs by von Rekowski
  \& Brandenburg (2003). Same model as in \Fig{Fig2}, but at $t\approx 158\,
  {\rm days}$.
  Collimation is due to boundary conditions.
  Right: simulations of outflows from accretion discs around black holes by
  Koide et al.\ (1998).
 }
 \label{Fig4}
\end{figure}

\section{Concluding remarks}

Our simulations show that a system consisting of a magnetically coupled star and
dynamo-active accretion disc can maintain
a clearly structured outflow. The outflow consists of a (hotter and denser)
slow stellar wind and a faster outer disc wind that are both pressure driven,
and a (cooler and less dense) fast magneto-centrifugally
accelerated inner disc wind within a conical shell (see \Fig{Fig4}, left panels).
In addition, in transition periods, when the magnetic star--disc connectivity changes,
there is a pressure driven (small azimuthal velocity and
angular momentum, the Alfv\'en surface close to the disc), hot and dense, but relatively fast
outflow between the stellar and disc winds.

Coexisting pressure and magneto-centrifugal driving mechanisms have also
been found by
Koide et al.\ (1998) who study the formation of relativistic jets
from accretion discs surrounding a black hole by performing numerical
simulations using a general relativistic magnetohydrodynamic code.
In their model, a global (initially vertical) magnetic field penetrates
the accretion disc which then leads to non-steady accretion and ejection
processes. According to their simulations, a jet is ejected from the close
vicinity of the black hole. The jet has a two-layered shell structure
consisting of a fast gas-pressure driven jet in the inner part and a slow
magnetically driven jet in the outer part, both of which are collimated
by the global poloidal magnetic field penetrating the disc (see \Fig{Fig4},
right panels). The fast jet
is due to a strong pressure increase due to shock formation at the inner edge
of the disc, caused by a fast accretion flow.

Our relatively fast pressure driven flow might be comparable with the fast jet.
However, contrary to the slow jet, our magneto-centrifugally accelerated
inner disc wind is even faster.


\acknowledgements
This work was supported by the PPARC Grant PPA/G/S/2000/00528.

\theendnotes

\end{opening}
\end{article}
\end{document}